\documentclass[12pt]{article}

\usepackage{amsmath,amssymb,amsthm,cite,mathtools}
\usepackage[margin=3cm]{geometry}

\usepackage{mathptmx,eucal}
\usepackage{bm} 

\title{Stora's fine notion of divergent amplitudes%
\thanks{To appear in Nuclear Physics B.}}

\author{Joseph C.~V\'arilly$^1$ and
Jos\'e M. Gracia-Bond\'ia$^{2,3}$ 
 \\[12pt]
{\footnotesize $^1$ Escuela de Matem\'atica,
Universidad de Costa Rica, San Jos\'e 11501, Costa Rica}\\[3pt]
{\footnotesize $^2$ Departamento de F\'isica Te\'orica,
Universidad de Zaragoza, Zaragoza 50009, Spain}\\[3pt]
{\footnotesize $^3$ Departamento de F\'isica,
Universidad de Costa Rica, San Pedro 11501, Costa Rica}\\[3pt]
}

\date{26 May 2016}

\DeclareMathOperator{\T}{T}         

\newcommand{\al}{\alpha}            
\newcommand{\bt}{\beta}             
\newcommand{\Dl}{\Delta}            
\newcommand{\dl}{\delta}            
\newcommand{\eps}{\varepsilon}      
\newcommand{\Ga}{\Gamma}            
\newcommand{\ga}{\gamma}            
\newcommand{\ka}{\kappa}            
\newcommand{\La}{\Lambda}           
\newcommand{\la}{\lambda}           
\newcommand{\sg}{\sigma}            

\newcommand{\bC}{\mathbb{C}}        
\newcommand{\bM}{{\mathbb{M}}}      
\newcommand{\bR}{{\mathbb{R}}}      
\newcommand{\bS}{\mathbb{S}}        

\newcommand{\del}{\partial}         
\newcommand{\otto}{\leftrightarrow} 
\newcommand{\7}{\dagger}            

\newcommand{\quarter}{\tfrac{1}{4}} 

\bmdefine{\lbold}{l}                

\newcommand{\vev}[1]{\langle\!\langle#1\rangle\!\rangle} 
\newcommand{\vevoo}[1]{\braCket{0}{#1}{0}} 
\newcommand{\word}[1]{\quad\text{#1}\quad} 

\def\wick:#1:{\,\mathopen:#1\mathclose:\,} 

\def\lLa^#1_#2{\Lambda^{#1}{}_{#2}}  
\def\ltLa_#1^#2{\Lambda_{#1}{}^{#2}} 

\newcommand{\braCket}[3]{\langle#1\mathbin|#2\mathbin|#3\rangle}

\makeatletter
\renewcommand{\section}{\@startsection{section}{1}{\z@}%
                       {-3.5ex \@plus -1ex \@minus -.2ex}%
                       {2.3ex \@plus.2ex}%
                       {\normalfont\large\bfseries}}
\makeatother

\begin{document}

\maketitle

\centerline{Raymond Stora, 1930--2015: \emph{in memoriam}}

\medskip

\begin{abstract}
\smallskip
Stora and coworkers refined the notion of divergent quantum amplitude,
somewhat upsetting the standard power-counting recipe. This
unexpectedly clears the way to new prototypes for free and interacting
field theories of bosons of any mass and spin.
\end{abstract}

\section{Exordium}
\label{sec:onthesurface}

One of us (JMG-B) learned of a flaw in the standard notion of
``superficially divergent amplitude'' from the lips of Raymond Stora,
quickly becoming aware of \textit{some} of the vistas opened by his
alternative notion during intense conversations at CERN in the winter
of 2013. In fairness, the notion should be attributed as well to
Nikolay M. Nikolov and Ivan Todorov, with whom Raymond was working at
the time on paper~\cite{NikolovST14}, wherein the matter is expounded
in convincing detail. We shall refer to the new notion of convergent
Feynman amplitude as the NST renormalization prescription.

\medskip

We begin by a review of causal Riesz distributions as introduced in
\cite{Bettina2}. This prelude smooths the way for the new notion of
divergent graph, valid for physical quantum fields (what ``physical''
means will be declared in due course). This helps to open the door to
the brave new world of string-local fields. Finally, in
Section~\ref{sec:gravedad} we show that, although homogeneity of the
amplitudes is lost, the concept in~\cite{NikolovST14} makes perfect
sense for massive theories.

\section{Causal Riesz distributions and massless field amplitudes}
\label{sec:thereis-acause-for-everything}
 
Let us invoke in somewhat simplified form a meromorphic family of
distributions on Minkowski space $\bM^4$ studied in~\cite{Bettina2}:
\begin{equation}
G(x;\al) 
:= \frac{e^{-i\pi\al}\Ga(-\al)}{4^{\al+2} \pi^2\Ga(\al + 2)}
(t^2 - r^2 - i0)^\al 
\equiv \frac{e^{-i\pi\al}\Ga(-\al)}{4^{\al+2}\pi^2\Ga(\al + 2)}
(x^2 - i0)^\al.
\label{eq:causal-Riesz} 
\end{equation}
The distribution $(x^2 - i0)^\al$ is well defined for 
$-2 < \Re\al < 0$; it can be extended analytically to non-integer
$\Re\al < -2$ by repeated applications of~$\square$; so
$(x^2 - i0)^\al$ can be regarded as meromorphic in~$\al$ with (simple)
poles at $-2 - n$ for $n = 0,1,2,\dots$. These are cancelled
in~\eqref{eq:causal-Riesz} by the poles of $\Ga(\al+2)$. The extension
prescription of analytic renormalization, obtained by discarding the
pole part in the Laurent expansion of $(x^2 - i0)^{\al+\eps}$, is
therefore straightforward whenever $\Re\al > - 2$, i.e., there is a
homogeneous extension. The relation
$$
\square G(x;\al) = G(x;\al - 1)
$$
holds, just as for the ordinary Riesz distributions. This is clear
from
$$
\square (x^2 - i0)^\al = 4\al(\al + 1) (x^2 - i0)^{\al-1},
$$
valid on the chosen domain, and then analytically extended. Note that
$iG(x;-1) = D_0^F(x)$, the Feynman propagator for massless scalars; so
$G(x;-l)$ for integer $l \geq 2$ is proportional to
$\square^{l-2}\dl(x)$. This is confirmed by a direct calculation of
the residues at $\al = -2,-3,\dots$.

The first aim of this paper is to investigate a generalization of all
this for massless particles of higher (integer) helicity. The quantum
Maxwell field can be built from the helicity $\pm 1$ massless unirreps
of the Poincar\'e group, under the form:
\begin{align}
F^{\mu\nu}(x) 
&:= i \sum_r \int d\mu(p)\, \bigl[ e^{i(px)} \bigl(
p^\mu e_r^\nu(p) - p^\nu e_r^\mu(p) \bigr) \, a_r^\7(p)
\notag \\
&\hspace{7em}
- e^{-i(px)} \bigl( p^\mu e_r^\nu(p)^* - p^\nu e_r^\mu(p)^* \bigr)
\, a_r(p) \bigr],
\label{eq:Maxwell-Proca-F-field} 
\end{align}
for appropriate creation operators $a_r^\7(p)$ and polarization
vectors $e_r^\nu(p)$. With $g_{\mu\nu}$ denoting the Minkowski metric
with $({+}{-}{-}{-})$ signature, routine computation establishes for
the vacuum expectation value of the two-point time-ordered
product~\cite{Stora05}:
\begin{align}
&\vev{\T F_{\mu\nu}(x)\, F_{\rho\sg}(x')} 
:= \vevoo{\T F_{\mu\nu}(x)\, F_{\rho\sg}(x')}
\notag \\
&= \bigl( g_{\mu\rho}\,\del_\nu\del_\sg
- g_{\nu\rho}\,\del_\mu\del_\sg - g_{\mu\sg}\,\del_\nu\del_\rho
+ g_{\nu\sg}\,\del_\mu\del_\rho \bigr) \, D_0^F(x - x')
=: f_{\mu\nu,\rho\sg}(\del) \, D_0^F(x - x')
\label{eq:TFF-massless} 
\end{align}
valid outside the diagonal $x = x'$. On the face of it, this
expression seems logarithmically divergent, since it homogeneously
scales like~$x^{-4}$; the field itself scales like~$x^{-2}$.

\medskip

For brevity, let us write $x^2 \equiv x^2 - i0$ hereinafter. In the
Epstein--Glaser program~\cite{EpsteinG73}, to renormalize a
distribution like $\vev{\T F_{\mu\nu}(x)\, F_{\rho\sg}(x')}$ in
position space is to find a suitable extension to the diagonal.
``Suitable'' means keeping the scaling behaviour of the original
distribution as much as possible. It also means satisfying physically
motivated and mathematically convenient requirements, in particular
Lorentz covariance and other symmetries.

Using translation invariance, extension of a distribution $f(x - x')$
to the diagonal is equivalent to extending $f(x)$, defined for
$x \neq 0$, to the origin in Minkowski space. Then the distribution
$x^{2\al} \equiv (x^2)^\al$ extends homogeneously for $\al > -2$; and
for integer~$\al \leq -2$, its extensions can be determined by the
complex-analytic methods in~\cite{NikolovST14} or the real-variable
methods in~\cite{Carme}, adopted in~\cite{Elara}. Thus for instance
the extensions of $x^{-4}$ are given by:
$$
R_4[x^{-4}]
= -\frac{1}{4} \square \biggl( x^{-2} \log\frac{x^2}{\ell^2} \biggr)
- i\pi^2\,\dl(x),
$$
with a length scale~$\ell$. This is log-homogeneous of bidegree
$(-4,1)$ in the terminology of~\cite{Elara}. (The Euclidean version is
$R_4[x^{-4}] = -\quarter \Dl(x^{-2}\log(x^2/\ell^2)) + \pi^2\,\dl(x)$;
the two cases differ only in the coefficient of~$\dl(x)$, arising from
the fundamental solutions of the Laplacian,
$\Dl(x^{-2}) = -4\pi^2\,\dl(x)$ in~$\bR^4$; and of the d'Alembertian,
$\square(x^{-2}) = 4i\pi^2\,\dl(x)$ in~$\bM^4$.)

For two-point functions which are polynomials in $x^{-2}$, these
procedures go a long way. For the sunset graph in massless $\phi_4^4$,
demanding Lorentz invariance, one can show \cite[(2.19)]{Elara} that
$$
R_4[x^{-6}]
= -\frac{1}{32} \square^2\biggl( x^{-2} \log\frac{x^2}{\ell^2} \biggr)
- \frac{5i\pi^2}{16} \,\square\dl(x),
$$
whose second term incidentally differs from the one in
\cite[Eq.~(5.29)]{NikolovST14} due to the precise usage of the
multiplicativity property of~\cite{Carme}.

One concludes that while unrenormalized two-point amplitudes are
\textit{homogeneous} functions for $x \neq x'$, they admit
\textit{log-homogeneous} extensions to the diagonal. The second index
in the bidegree indicates the power of the logarithm, counting the
number of successive extensions for distributions presenting
subdivergences, in general: the sunset graph is quadratically
divergent, but still primitive in this dispensation. The matter was
treated in detail for many graphs of the massless $\phi_4^4$ theory
in~\cite{Elara}, albeit in the Euclidean signature; happily, only
minor modifications are needed for the Minkowskian version. There has
been a crop of relatively recent papers dealing with this kind of
problem~\cite{NikolovST14,Elara,DuetschFKR}, reaching similar
conclusions.

\medskip

Things appear to be more complicated when the unrenormalized amplitude
has an angular dependence, as in our present
case~\eqref{eq:TFF-massless}. Since $\del_\mu\del_\rho(x^{-2}) 
= -2(g_{\mu\rho} x^2 - 4 x_\mu x_\rho)x^{-6}$, we compute (for
$x \neq 0$):
\begin{align}
& \bigl( g_{\mu\rho}\,\del_\nu\del_\sg
- g_{\nu\rho}\,\del_\mu\del_\sg - g_{\mu\sg}\,\del_\nu\del_\rho
+ g_{\nu\sg}\,\del_\mu\del_\rho \bigr) [x^{-2}] 
\notag \\
&\quad = -4 \Bigl( (g_{\mu\rho}\,g_{\nu\sg} - g_{\nu\rho}\,g_{\mu\sg}) x^2
- 2(g_{\mu\rho}\,x_\nu x_\sg - g_{\nu\rho}\,x_\mu x_\sg
- g_{\mu\sg}\,x_\nu x_\rho + g_{\nu\sg}\,x_\mu x_\rho) \Bigr) x^{-6}
\notag \\
&\quad =: h_{\mu\nu,\rho\sg}(x)\, x^{-6},
\label{eq:peace-and-harmony} 
\end{align}
where each $h_{\mu\nu,\rho\sg}(x)$ is a homogeneous quadratic 
polynomial. 

In fact, each of these polynomials is \textit{harmonic} in the
Minkowskian sense. To see that, it is enough to apply
$\square(x^2) = 8$ and $\square(x_\mu x_\nu) = 2 g_{\mu\nu}$ to the
quadratic polynomial in~\eqref{eq:peace-and-harmony}, to get
$$
\square h_{\mu\nu,\rho\sg}(x) 
= -4(8 - 8)(g_{\mu\rho}\,g_{\nu\sg} - g_{\nu\rho}\,g_{\mu\sg}) = 0.
$$

Actually, these $h_{\mu\nu,\rho\sg}$ form a 
\textit{basis} for the vector space of quadratic harmonic polynomials
on~$\bM^4$. Due to (skew)symmetry under the exchanges 
$\mu \otto \nu$ and $\rho \otto \sg$, and symmetry under 
$(\mu,\nu) \otto (\rho,\sg)$ and $(\mu,\nu) \otto (\sg,\rho)$, there
are $9$ linearly independent $h_{\mu\nu,\rho\sg}$; whereas the
harmonic homogeneous polynomials of degree~$k$ on~$\bM^4$ (or on
$\bR^4$, for that matter) form a space of dimension $(k+1)^2$
\cite[Sect.~9.3]{AndrewsAR99}.

\section{The NST renormalization prescription}
\label{sec:levedad}

The task then becomes to extend to the origin functions of the form
$x^{2\al}\,H_k(x)$, where $H_k$ is a homogeneous polynomial of
degree~$k$ that is also (Minkowskian) harmonic.  There are two reasons
to hope that the ``radial'' extensions of~\cite{NikolovST14, Elara}
may prove equal to the task.  The first is the off-origin calculation:
\begin{align}
\square(x^{2\al}\,H_k(x)) &= \square(x^{2\al}) H_k(x) 
+ 2 \del^\mu(x^{2\al}) \,\del_\mu(H_k(x)) + x^{2\al}\,\square(H_k(x))
\notag \\
&= 4\al(\al + 1) x^{2\al-2} H_k(x) 
+ 4\al x^{2\al-2} x^\mu\,\del_\mu(H_k(x))
\notag \\
&= 4\al(\al + k + 1) x^{2\al-2} H_k(x),
\label{eq:push-back} 
\end{align}
where we have used harmonicity: $\square H_k = 0$, and homogeneity:
$x^\mu\,\del_\mu H_k = kH_k$. These relations show that the family of
$x^{2\al}H_k(x)$ also act like the causal Riesz
distributions~\eqref{eq:causal-Riesz}; a suitable normalization is
$$
G(x;\al,k) 
:= \frac{e^{-i\pi\al}\Ga(-\al)}{4^{\al+2}\pi^2\,\Ga(\al+k+2)}\,
x^{2\al} H_k(x);
$$
and from \eqref{eq:push-back} we get at once:
\begin{equation}
\square\, G(x;\al,k) = G(x; \al - 1, k).
\label{eq:pushed-around} 
\end{equation}
The extension prescription of analytic renormalization now tells us
that there is a \textit{homogeneous} extension whenever 
$\al > - k - 2$. In particular, the case of
interest~\eqref{eq:peace-and-harmony} has $\al = -3$ and $k = 2$.
Since $-3 > -4$, the na\"ive power-counting recipe is overridden: the
time-ordered product \eqref{eq:TFF-massless} does extend
homogeneously to the origin, the result being none other than:
$$
\vev{\T F_{\mu\nu}(x)\, F_{\rho\sg}(x')} 
= \frac{i}{4\,\pi^2}\, f_{\mu\nu,\rho\sg}(\del)
\frac{1}{(x - x')^2 - i0}\,,
$$
as many a physicist, taking a cue from the commutation relations
\cite[Aufgabe~7.5]{GreinerR93}, would have written at the outset. In
other words, the apparent singularity was removable; according to the
lore of renormalization of massless amplitudes, truly renormalization
has not taken place.

\medskip

The general criterion~\cite[Corl.~5.4]{NikolovST14} is: a two-point
unrenormalized Feynman amplitude in Minkowski space of the form
$h_k(x)/(x^2 \pm i0)^s$ for $x \neq 0$ has an homogeneous extension if
and only if its ``degree of harmonicity'' $k$ is greater than the
``degree of divergence'' $2s - k - 4$. 

Furthermore, in this case the homogeneous extension is unique if we
impose Lorentz covariance. This needs to be properly understood. Once
a homogeneous extension of $x^{2\al} H_k(x)$ has been found, any other
such extension can differ from it only by a distribution
$P(\del)\,\dl(x)$ supported at the origin, where $P(x)$ is a
homogeneous polynomial of degree $2(-\al) - k - 4$, the superficial
degree of divergence. In our example, this degree is~$0$, so $P(x)$
would be a constant. However, $H_k(x)$ is not constant: indeed, it
transforms under a representation of the Lorentz group on the space of
harmonic homogeneous polynomials of degree~$k$, and $P(x)$ must
transform likewise. The upshot is that $P(x)$ must be at least
divisible by such a harmonic homogeneous polynomial, so that
$\deg P \geq k$. Thus, again the condition $k > 2(-\al) - k - 4$ [in
the example: $2 > 0$] is enough to ensure that the Lorentz-covariant
extension of $x^{2\al} H_k(x)$ is unique. In fine: the off-diagonal
function \eqref{eq:TFF-massless} extends to a Lorentz-covariant
time-ordered product, without ambiguity. Equivalently, one can argue
in the spirit of the \textit{on-shell extension} of amplitudes by
Bahns and Wrochna~\cite{BahnsW14}: the decisive fact is that the
differential equation~\eqref{eq:pushed-around} is extended to the
origin, too.

\section{The prescription for higher helicities \dots}
\label{sec:friend-come-up-higher}

Similarly to the above, there is a free quantum field
$R_{\al\bt\rho\tau}(x)$, the linearized Riemann tensor, corresponding
to helicity-$2$ particles and transforming as a rank~$4$ tensor, with
the symmetry properties:
$$
R_{\al\bt\ka\tau}(x) = - R_{\bt\al\ka\tau}(x) = - R_{\al\bt\tau\ka}(x)
= R_{\ka\tau\al\bt}(x).
$$
One analogously finds for this:
\begin{align}
\vev{\T R_{\al\bt\ka\tau}(x)\, R_{\rho\sg\la\ga}(x')} 
&= \sum \pm G_{\bt\tau,\sg\ga} \,\del_\al\del_\ka\del_\rho\del_\la
\,D_0^F(x - x') + 15 \text{ similar terms}
\notag \\
&=: \frac{16\pi^8}{3} h_{\al\bt\ka\tau,\rho\sg\la\ga}(x)\,
D_0^F(x - x')^5;
\label{eq:cross-roads} 
\end{align}
where $G_{\bt\tau,\sg\ga} := \frac{1}{8} \bigl( g_{\bt\sg} g_{\tau\ga}
+ g_{\bt\ga} g_{\tau\sg} - g_{\bt\tau} g_{\sg\ga} \bigr)$ and the
``similar terms'' are obtained by permuting the indices under exchange
of $(\al,\bt,\rho,\sg)$ with $(\ka,\tau,\la,\ga)$,
$(\tau,\ka,\la,\ga)$, $(\ka,\tau,\ga,\la)$ and $(\tau,\ka,\ga,\la)$
respectively; the signs are those that respect the aforementioned
symmetries.%
\footnote{The expression for $G_{\bt\tau,\sg\ga}$ appears in the
graviton propagator, see for instance \cite[Eq.~1.77]{Hamber09}.}
Therefore, $h_{\al\bt\ka\tau,\rho\sg\la\ga}(x) 
= \sum \pm G_{\bt\tau,\sg\ga}\, q_{\al\ka\rho\la}(x)$ is likewise a
sum of $16$ quartic \textit{harmonic} polynomials, coming from
$\del_\al \del_\ka \del_\rho \del_\la [x^{-2}] 
=: q_{\al\ka\rho\la}(x)\,x^{-10}$ by direct calculation, such as:
\begin{align*}
q_{\al\ka\rho\la}(x) 
&:= 48 x_\al x_\ka x_\rho x_\la 
+ \bigl( g_{\al\ka} g_{\rho\la} + g_{\al\la} g_{\ka\rho}
+ g_{\al\rho} g_{\ka\la}\bigr)\, x^4
\\
&\quad - 6 \bigl( g_{\al\ka} x_\rho x_\la + g_{\al\rho} x_\ka x_\la
+ g_{\al\la} x_\ka x_\rho + g_{\ka\rho} x_\al x_\la 
+ g_{\ka\la} x_\al x_\rho + g_{\rho\la} x_\al x_\ka \bigr)\, x^2.
\end{align*}
The harmonic property $\square q_{\al\ka\rho\la}(x) = 0$ is easily
checked directly, using:
\begin{gather*}
\square(x^4) = 24 x^2,  \qquad
\square(x_\rho x_\la x^2) = 2 g_{\rho\la} x^2 + 16 x_\rho x_\la\,,
\\
\square(x_\al x_\ka x_\rho x_\la)
= 2 g_{\al\ka} x_\rho x_\la + 5 \text{ similar terms}.
\end{gather*}

Just as before, these $h_{\al\bt\ka\tau,\rho\sg\la\ga}$ constitute a
basis of the $25$-dimensional space of quartic homogeneous harmonic
polynomials on~$\bM^4$. Indeed, taking into account the $20$
independent components of $R_{\al\bt\rho\tau}(x)$ and the four
mentioned symmetries of the cross-indexes, the number of independent
$h_\bullet$-polynomials in this case is $(20)^2/2^4 = 25$.

Now, on the face of it there is a quadratic divergence here -- the
field scales like~$x^{-3}$. However, since $4 > 10 - 4 - 4$, by the
same token as above, the finer NST criterion shows that the vacuum
expectation value of the time-ordered $2$-point function for the
$R$-tensor field is a \textit{convergent} amplitude.

\medskip

How to generalize to higher integer helicities should be clear now:
among the free point-local fields for helicity~$h$ there are two
tensor fields with apparently optimal ultraviolet behaviour in
relative terms, namely, they scale~like $x^{-h-1}$: the \textit{field
strength} $F_{\mu_1\nu_1,\dots,\mu_h\nu_h}$ of rank~$2h$, symmetric
under exchange of any of the pairs $(\mu_i,\nu_i) \otto (\mu_j,\nu_j)$
and skewsymmetric under exchange inside the pairs; and its potential
$A_{\mu_1,\dots,\mu_h}$ of rank~$h$, which is totally
symmetric~\cite{MundDO16,Weinberg95I}. The quantum fields associated
to the representation $(h,0) \oplus (0,h)$ are ``physical'' in that
their classical counterparts are measurable.

``Apparently'' we say, because in fact 
$\vev{\T F_{\mu_1\nu_1,\dots,\mu_h\nu_h}
F_{\al_1\bt_1,\dots,\al_h\bt_h}}$ is a convergent amplitude, as we
have seen for $h = 1,2$. Whereas the $2$-point function for the
potentials carries a problematic existence, due to gauge freedom (or
slavery) and the impossibility, starting with the photon, for
$A_{\mu_1,\dots,\mu_h}$ to live on Hilbert space.

\section{\dots\ and its consequence: a gauge-free world?}
\label{sec:tied-up-in-strings}

By abandoning point-localization, it is feasible to construct
$A$-fields for any boson particle that share in the good ultraviolet
properties of the field strengths. This fact has been known for over
ten years now~\cite{MundSY04,MundSY06}, and has the potentiality to
drastically change the game of perturbative quantum field theory.

The field strengths remain pointlike. To keep notations simple, here
we just exhibit a (lightlike) string-local potential field for the
photon:
$$
A^\mu(x,l) := \int_0^\infty dt\, F^{\mu\nu}(x + tl) \,l_\nu \,,
$$
with $l = (l^0,\lbold)$ a null vector. The definition depends only on
the ray of~$l$, which is a point of the celestial sphere~$\bS^2$, or
the light front uniquely associated to it.

A comment is in order here. Previous formulations of string-local
fields were based on modular localization theory, which naturally
suggests the use of spacelike strings~\cite{Schroer14}. However, in
interacting models this leads to almost intractable complications at
third order of perturbation theory. For purely massive models, there
is a huge advantage in employing null strings, since then the field is
actually a well-behaved function on the $l$-variable, not just a
distribution like in the spacelike case. In models containing massless
particles, use of null strings generates a \textit{sui generis}
ultraviolet-infrared problem, which needs to be and can be dealt with
by appropriate recipes. Note that all null directions are on the same
footing: each one carries its own cyclic subspace, and these are
shuffled around by the Lorentz transformations -- see right below.

The operator-valued distribution $A$ ``lives'' on the same Fock space
as~$F$, and its main properties are the following:
\begin{itemize}
\item
Transversality: $\bigl(l\, A(x,l)\bigr) = 0$.
\item
Pointlike differential:
$\del^\mu A^\nu(x,l) - \del^\nu A^\mu(x,l) = F^{\mu\nu}(x)$.
\item
Covariance: let $U$ denote the lifting (or ``second quantization'') of
Wigner's unirrep of the Poincar\'e group on the one-particle states.
Then
$$
U(a,\La) A^\mu(x,l) U^\7(a,\La) 
= A^\nu(\La x + a, \La l) \,\ltLa_\nu^\mu
= (\La^{-1})_{\;\;\nu}^\mu A^\nu(\La x + a, \La l).
$$
\item
Locality: 
$[A_\mu(x,l), A_\nu(x',l')] = 0$ when the strings $x + tl$ and
$x' + t'l'$ are causally disjoint.
\end{itemize}

The very concept of gauge disappears, since this potential
\textit{vector}, with all the good properties, is uniquely defined.
The formalism appears more exotic than the usual one, in that a new
variable is invoked. ``The choice of what kind of field describes an
observed particle is really a matter of choice: try what type of field
describes best the observed data''~\cite{Veltman94}. It is however
more mundane, in that it allows us to remain in physical Hilbert
spaces: the ghosts can depart, since there is need for them no
longer.

Of course, the string ``ought not to be seen'', and the program
becomes to demonstrate whether, and how, this simple criterion is
enough to determine interaction vertices and govern perturbative
renormalization of string-local models of so-called (Abelian and
non-Abelian) gauge interactions~\cite{tHooft16} from the Lie algebra
structure, down to every relevant detail~\cite{Rosalind,Helene}. This 
includes models with massive intermediate vector bosons -- see the 
following section.

The above construction works in a parallel way for \textit{all} the
other integer-helicity cases, like linear gravity, which now are
gauge-free, and seen to possess the same ultraviolet properties as
scalar particles.%
\footnote{%
\label{fn:tres-reyes}%
It appears tempting to redo some of the graviton-scattering
calculations in~\cite{AlvarezGMM16}, performed in the framework of
unimodular gravity, using the $A(x,l)$-field companion of the
linearized Riemann tensor.}
What we realize is that the construction of string-local
fields~\cite{MundSY04,MundSY06} rests on the bedrock of a
never-ambiguous time-ordered product of the field strengths.

\section{Massive field amplitudes}
\label{sec:gravedad}

With a suitable change of the polarization vielbeins $e_r^\nu$, the
very formula \eqref{eq:Maxwell-Proca-F-field} describes a
skewsymmetric quantum field for massive spin~$1$
particles~\cite{Stora05}. In the massive case,
Eq.~\eqref{eq:TFF-massless} holds as well. A small miracle is
involved here, since
$$
F_{\mu\nu}(x) = \del_\mu B_\nu(x) - \del_\nu B_\mu(x),
$$
where $B$ denotes the Proca field, and for it, outside the diagonal
$x = x'$:
$$
\vev{\T B_\mu(x) B_\nu(x')}
= i(g_{\mu\nu} + \del_\mu\del_\nu/m^2)\, D^F(x - x'),
$$
with just $D^F$ denoting the massive scalar Feynman propagator. Thus
one would expect fourth-order derivatives (a quadratic divergence) in
$\vev{\T FF'}$. But they all cancel, so the $2$-point time-ordered
function off the diagonal $x = x'$ looks exactly like the one
in~\eqref{eq:TFF-massless}:
\begin{align}
\vev{\T F_{\mu\nu}(x)\, F_{\rho\sg}(x')} 
= \bigl( g_{\mu\rho}\,\del_\nu\del_\sg - g_{\nu\rho}\,\del_\mu\del_\sg
- g_{\mu\sg}\,\del_\nu\del_\rho + g_{\nu\sg}\,\del_\mu\del_\rho \bigr)
\, D^F(x - x'),
\label{eq:TFF-massive} 
\end{align}
but with the massive propagator replacing the massless one.

That still looks logarithmically divergent. However, since the
ultraviolet properties in both cases are the same, most physicists
would conclude without hesitation that the formula makes sense and
extends $\vev{\T F_{\mu\nu}(x)\, F_{\rho\sg}(x')}$ to the diagonal. We
cite Todorov in this context: ``Introducing \dots\ masses in the
analysis of small distance behaviour seems to be just adding technical
details to the general picture''~\cite{Todorov16}.

The conclusion is correct, and can be substantiated in at least two
rather different ways.
\begin{itemize}

\item
We recall the expansion of $D^F$ in the vicinity of $m = 0$:
\begin{equation}
D^F(x) = D^F_0(x) + m^2 \bigl[ f_1(m^2 x^2) \log(-m^2(x^2 - i0))
+ f_2(m^2 x^2) \bigr],
\label{eq:how-mass-matters} 
\end{equation}
where $f_1$, $f_2$ are analytic. In \cite[Sect.~6]{Duetsch15}, it is
shown that the basic postulate of Epstein--Glaser renormalization, to
wit, that the renormalized amplitudes scale like the unrenormalized
ones, up to logarithmic corrections, can be strengthened, in that
these corrections -- albeit necessarily introducing a new mass scale
-- do not change the dependence on~$m$ in~\eqref{eq:how-mass-matters};
so \eqref{eq:TFF-massive} extends to the diagonal without further ado.

\item
A method in the spirit of the present paper is as follows
\cite{Bettina1}.%
\footnote{This is actually the same paper as \cite{Bettina2}, but in
the published version the pertinent section was withdrawn, because the
referee could not make head or tail of~it.}

We can modify $G(x;\al)$ in~\eqref{eq:causal-Riesz} by extracting the
finite part of $\Ga(-\al) x^{2\al}$ for $\al = 0,1,2,\dots$. This is
equivalent to renormalizing the convolution powers of the massless
Feynman propagator; these are all primitives, which means that only
the first power of the logarithm appears in:
\begin{align*}
F(x;\al) &:= G(x;\al) \word{for} \al \neq 0,1,\dots;
\\
F(x;n) &:= \frac{e^{-i\pi n}\,x^{2n}}{4^{n+2}\pi^2\,n!(n + 1)!}
\Bigl( \log\frac{m^2x^2}{4} - \psi(n+2) - \psi(n+1) - i\pi \Bigr),
\\
&\qquad
\text{for $n = 0,1,\dots$; where $\psi$ is the digamma function}.
\end{align*}
Note the choice $m = 1/l$ here.
\end{itemize}

Now $\square F(x;\al) = F(x;\al - 1)$ holds without
restriction~\cite{BolliniGGD64}, so in fact we may write
$$
F(x;\al) = -i\,\square^{-1-\al} D^F_0(x),
$$
for all $\al \in \bC$, and we have a perfect generalization of Riesz
theory. Moreover, the series $\sum_{n=-1}^\infty m^{2n+2} F(x,n)$
solves the massive Klein--Gordon equation with the convolution unit as
source \cite{AbramowitzS65,Schnetz97}:
$$
\sum_{n=-1}^\infty m^{2n+2} F(x,n)
= -im \frac{K_1(m\sqrt{-x^2}\,)}{4\pi^2\sqrt{-x^2}} = D^F(x).
$$
So let us define, for $H_k$ homogeneous harmonic of order~$k$:
\begin{align*}
F(x;\al,k) &= G(x;\al,k) \word{for} \al \neq 0,1,\dots;
\\
F(x;n,k) &:= H_k(x)\, 
\frac{e^{-i\pi n}\,x^{2n}}{4^{n+2}\pi^2\,n!(n + k + 1)!}
\biggl( \log\frac{m^2x^2}{4} - \psi(n+2) - \psi(n+1) - i\pi \biggr).
\end{align*}
Finally, it is clear that the formula
$$
\vev{\T F_{\mu\nu}(x)\, F_{\rho\sg}(x')} 
= f_{\mu\nu,\rho\sg}(\del)\,D^F(x - x'),
$$
valid for $x \neq x'$, extends to the diagonal without further
renormalization being necessary.

What about higher spins? Following~\cite{MundDO16}, we compute the
expected value of the time-ordered product of the linearized Riemann
tensor for \textit{massive} gravitons, with a result identical
to~\eqref{eq:cross-roads}, except that instead of $G_{\bt\tau,\sg\ga}$
as in Sect.~\ref{sec:friend-come-up-higher}, one finds
$\frac{1}{8} \bigl( g_{\bt\sg }g_{\tau\ga} + g_{\bt\ga} g_{\tau\sg}
- \frac{2}{3} g_{\bt\tau} g_{\sg\ga}\bigr)$.%
\footnote{A similar expression with the $\frac{2}{3}$ coefficient
appears in the massive graviton propagator given in
\cite[Sect.~1.5]{Zee10}.}
This difference between the massive and the massless cases is
immaterial for harmonicity since, as we remarked earlier, the
polynomials $q_{\al\ka\rho\la}$ are already harmonic. Therefore
$\vev{\T R_{\al\bt\ka\tau}(x)\, R_{\rho\sg\la\ga}(x')}$ extends to the
diagonal, without further ado.

We conjecture that our conclusions extend to all the massive
$F_{\mu_1\nu_1,\dots,\mu_h\nu_h}$-fields.

\section{Conclusion}
\label{sec:coda}

Two small miracles do not a big miracle make. Nevertheless, it is
surprising and gratifying that, against appearances, for massive or
massless particles of respectively integer spin or helicity~$j$, the
quantum fields associated to the representation $(j,0) \oplus (0,j)$
enjoy the same optimal UV properties. These are inherited by the
string-local true tensor fields $A_{\mu_1,\dots,\mu_h}(x,l)$
constructed from them.%
\footnote{As we were readying this paper for publication, we were made
aware of the article~\cite{Nikolov16}. It also seeks to transfer
results from massless to massive models, in a direction different
from~ours.}

\section*{Acknowledgements}

We are most grateful to Jens Mund for many discussions on string-local
fields, raising some of the issues discussed here, and for making
Ref.~\citen{MundDO16} available to us. Thanks are due to Michael
D\"utsch for timely comments on the manuscript, to Carmelo P. Mart\'in
for comments translating into our footnote~\ref{fn:tres-reyes}, to
Bert Schroer for constant prodding on the subject of string-local
fields and to Ivan Todorov for bringing reference~\cite{Nikolov16} to
our attention. The referee's comments and questions were very
instrumental in improving the paper. This research was generously
helped by the program ``Research in Pairs'' of the Mathematisches
Forschungsinstitut Oberwolfach in November~2015. The project has
received funding from the European Union's Horizon~2020 research and
innovation programme under the Marie Sk{\l}odowska-Curie grant
agreement No.~690575, and from the COST Association through the COST
Action QSPACE MP1405. JCV acknowledges support from the
Vicerrector\'ia de Investigaci\'on of the Universidad de Costa~Rica.

\end{document}